\documentclass[a4paper]{jpconf}

\newcommand{\p}[1]{(\ref{#1})}

\newcommand{\be}{\begin{equation}}
\newcommand{\ee}{\end{equation}}
\newcommand{\bea}{\begin{eqnarray}}
\newcommand{\eea}{\end{eqnarray}}

\newcommand{\ba}{\begin{array}} \newcommand{\ea}{\end{array}}

\newcommand{\nn}{\nonumber}

\usepackage{amscd,amsmath,amssymb}

\begin{document}
\title{Oscillators from nonlinear realizations}
\author{N Kozyrev and S Krivonos}
\address{Bogoliubov  Laboratory for Theoretical Physics, JINR,
141980 Dubna, Russia}
\ead{nkozyrev,krivonos@theor.jinr.ru}

\begin{abstract}\noindent
We construct the systems of the harmonic and Pais-Uhlenbeck oscillators, which are invariant with respect to arbitrary noncompact Lie algebras. The equations of motion of these systems can be obtained with the help of the formalism of nonlinear realizations. We prove that it is always possible to choose time and the fields within this formalism in such a way that the equations of motion become linear and, therefore, reduce to ones of ordinary harmonic and Pais-Uhlenbeck oscillators. The first-order actions, that produce these equations, can also be provided. As particular examples of this construction, we discuss the $so(2,3)$ and $G_{2(2)}$ algebras.
\end{abstract}

\setcounter{page}{1}
\setcounter{equation}{0}

\section{Introduction}
In the papers \cite{KN1} and \cite{KLS2}, the dynamical systems, which provide nonlinear realizations [3-6] of the $su(1,2)$, $so(2,3)$ and $G_{2(2)}$ algebras, were considered. The equations of motion of these systems were obtained within the coset approach as the conditions on some of the Cartan forms of the specifically chosen coset spaces. Other forms of these cosets were used to construct the invariant actions. The equations of motion, obtained in \cite{KN1}, \cite{KLS2}, can be viewed as deformations of the equations of motion of usual harmonic oscillators. The ability to obtain the oscillator equations of motion is strongly related to the fact that these algebras admit the 5-graded decomposition. And, due to the fact that at least one real form of any simple Lie algebra, aside of $sl(2)$, admits this decomposition \cite{BG}, it is possible to realize all these algebras in terms of time and fields, that satisfy the deformed oscillator equations of motion.

The idea of the papers \cite{KN1}, \cite{KLS2} can be traced back to the paper \cite{KLS1}, where the deformations of the $\ell$-conformal Galilei algebras were considered and it was noted that the deformed commutation relations resemble ones of a very general class of Lie algebras. In this sense, the appearance of the oscillators is not surprising, as for the $\ell$-conformal Galilei algebras the nonlinear realizations that produce equations of motion of the harmonic and Pais-Uhlenbeck oscillators, are known \cite{Galaj1}, \cite{AGGM}.

An interesting feature of the construction of the invariant oscillators in \cite{KLS2} was that the second-order equations of motion of the $so(2,3)$ invariant system were still linear, though both $so(2,3)$ and $G_{2(2)}$-invariant actions were deformed, and required an introduction of some additional semi-dynamical variables. Moreover, in \cite{KN1} the redefinition of the coordinates and the momenta was found, which brought the equations of motion and the Hamiltonian of the $su(1,2)$ deformed oscillators to the standard form. As, obviously, the exact form of the equations of motion heavily depends on the parametrization of the coset space, one may ask the question whether these equations can be made linear by the change of variables.

The purpose of this paper is to prove that for any simple Lie algebra it is possible to construct parametrization, which ensures linearity of equations of motion. This parametrization can be called homogeneous, as it is just required to place all the dynamical fields together with their generators into a single exponent. We also consider a general recipe for construction of the action and analyze how it produces the expected oscillator equations of motion. Then we consider, for illustrative purposes, the example of $so(2,3)$ in homogeneous parametrization and find another (``matryoshka'') parametrization of the $G_{2(2)}$ coset, which also produces linear equations of motion.

\section{General construction}
\subsection{Linear equations of motion}
As was already noted, we consider algebras with the structure \cite{KLS1}, \cite{KLS2}
\bea\label{lconfG}
&&\big[ L_{m},L_{n} \big]= (m-n)L_{m+n}, \; m,n=-1,0,1;  \;\; [M_a, M_b ] \sim M_c, \nn \\
&&\big[ L_{m}, G_{r,\mu} \big] = \big({\ell} m -r\big)G_{r+m,\mu}, r=-\ell,\ldots,\ell; \; \; [M, G] \sim G,\\
&&\big[G_{r,\mu}, G_{s,\nu}\big] = \delta_{r+s,0}f^a(r,\mu,\nu) M_a  + g^m (r,s,\mu,\nu) L_m,  \nn
\eea
where $\ell$ is a half-integer to have the finite number of $G_{r,\mu}$ and $L_m$ form the conformal algebra in one dimension. This structure of the algebra is called graded, as all generators can be classified with respect to the action of $L_0$. The grade of the algebra \p{lconfG} depends on $\ell$. It was proven in \cite{BG} that a 5-graded structure, which corresponds to $\ell=\frac{1}{2}$, exists for at least one real form of any simple Lie algebra, though different graded structures are known for various algebras. In this paper, we concentrate on the case $\ell=\frac{1}{2}$.

The equations that are deformed equations of motion of harmonic oscillators can be obtained as conditions $\Omega^{r,\mu}=0$ on the $G_{r,\mu}$ Cartan forms of the coset spaces
\be\label{simplecoset}
g = e^{t\big( L_{-1} + \omega^2 L_1  \big)} \,\times\, \mbox{exponents of } u^{r,\mu}G_{r,\mu}.
\ee
There are a lot of coset spaces with this structure, which differ by exact arrangement of $G_{r,\mu}$ generators. However, it is always possible, by redefining $t$ and $u^{r,\mu}$, to bring all the $G_{r,\mu}$ inside a single exponent. Let us show that in this case the resulting equations of motion are necessarily linear and for $\ell=\frac{1}{2}$ coincide with those of harmonic oscillators of identical frequency. Indeed, using standard techniques, one may calculate the Cartan forms of the coset element
\bea\label{homcoset}
g &=& e^{t\big( L_{-1} + \omega^2 L_1  \big)}e^{u \cdot G}, \; u\cdot G = \sum_{r,\mu} u^{r,\mu} G_{r,\mu} \; \Rightarrow \nn \\\label{homforms}
g^{-1}dg &=& e^{-u \cdot G} dt\big( L_{-1} + \omega^2 L_1\big) e^{u \cdot G} +  e^{-u \cdot G}d e^{u \cdot G} = \nn \\
&=& e^{-u \cdot G}\wedge dt\big( L_{-1} + \omega^2 L_1\big) +  \frac{1 - e^{-u \cdot G}}{u \cdot G} \wedge\big( du \cdot G \big).
\eea
Here $\wedge$ is the Zumino notation for the nested commutators: $X\wedge Y = [X,Y]$, $X^2\wedge Y= [X,[X,Y]]$, e.t.c.

Rewriting each of the exponents as a sum of even and odd part $e^X = \cosh X + \sinh X$,
\bea\label{exps}
&&e^{-X}\wedge Y = \cosh X \wedge Y   - \sinh X\wedge Y = \cosh X \wedge Y - \frac{\sinh X}{X}\wedge [X,Y], \nn \\
&&\frac{1-e^{-X}}{X} \wedge dX = \frac{1 - \cosh X}{X^2} \wedge [X,dX] + \frac{\sinh X}{X} \wedge dX,
\eea
the Cartan form can be represented as
\bea\label{homformsrewr}
\Omega = dt \big(L_{-1} + \omega^2 L_1\big) - \frac{\cosh{(u\cdot G)}-1}{(u\cdot G)^2}\wedge \big[u\cdot G, du\cdot G -dt \big[u\cdot G, L_{-1} + \omega^2 L_1 \big]\big] +\nn \\
+\frac{\sinh (u \cdot G)}{(u\cdot G)}\wedge\big( du\cdot G -dt \big[u\cdot G, L_{-1} + \omega^2 L_1 \big]  \big).
\eea
As the structure of the algebra \p{lconfG} suggests, the expression $du\cdot G -dt \big[u\cdot G, L_{-1} + \omega^2 L_1 \big]$ is composed only of $G$ generators. Moreover, calculating $N$ nested commutators of this expression with $u \cdot G$, one may find that the result is proportional to $G$ again if $N$ is even but contains only $L$ and $M$ generators if $N$ is odd. Therefore, one may conclude that $G_{r,\mu}$  appear only in the last line of \p{homformsrewr} and
\be\label{Gforms}
\sum_{r,\mu}\Omega^{r,\mu} G_{r,\mu} = \frac{\sinh (u \cdot G)}{(u\cdot G)}\wedge\big( du\cdot G -dt \big[u\cdot G, L_{-1} + \omega^2 L_1 \big]  \big).
\ee
As the $\sinh(x)/x$ begins with $1$, its adjoint action is always invertible and the equations, which follow from $\Omega^{r,\mu}=0$, are equivalent to
\be\label{geneom}
\dot{u}\cdot G - \big[u\cdot G, L_{-1} + \omega^2 L_1 \big] =0.
\ee
For $\ell =\frac{1}{2}$, $r$ takes only values $-\frac{1}{2},\frac{1}{2}$, and it is not difficult to evaluate the expression \p{geneom} explicitly. It would result in the equations of motion of a few (depending on the representation of $M$) harmonic oscillators
\be\label{osceom}
{\dot u}^{-1/2,\mu} + u^{1/2,\mu}=0, \; \; {\dot u}^{1/2,\mu} -\omega^2 u^{-1/2,\mu}=0 \; \Rightarrow \; {\ddot u}^{-1/2,\mu} +\omega^2 u^{-1/2,\mu} =0.
\ee
If there exists a basis for the algebra with different $\ell$, one may obtain the Pais-Uhlenbeck oscillators.

Equations \p{geneom} explicitly linear and, therefore, similar to ones obtainable from the Galilei algebras. Let us also note that $M_a$ generators appear only in the second term in the first line of \p{homformsrewr}, and their forms
\be\label{Mforms}
\sum_{a}\Omega^{a} M_{a} = - \frac{\cosh{(u\cdot G)}-1}{(u\cdot G)^2}\wedge \big[u\cdot G, du\cdot G -dt \big[u\cdot G, L_{-1} + \omega^2 L_1 \big]\big] |_{M},
\ee
where $|_{M}$ implies that we consider only $M_a$ generators, also vanish on equations of motion.

\subsection{The action}
The Cartan forms of $G_{r,\mu}$ generators were already used in the derivation of the equations of motion, and the only ingredients to construct the action appear to be the forms of $M_a$. Let us note, however, that due to the general transformation law of the Cartan forms
\be\label{formtr}
 g_0 g = g^\prime h \; \Rightarrow\; \Omega^\prime = (g^\prime)^{-1} dg^\prime = h\Omega h^{-1} - dh h^{-1},
\ee
the forms of the small subgroup transform non-homogeneously. This is not a problem if the small subgroup is a product of $U(1)$'s, as the $dh h^{-1}$ term in this case will be full differential and its integral would vanish, making integral of the whole form invariant. While it is not possible to use this directly in the case of a non-Abelian small subgroup, one may still split its generators $\big\{M_a \big\} \rightarrow\big\{ M_0=u(1), M_A\big\}$, move the $M_A$ to the coset space with parameters $\Lambda^A$ and use the integral of the form of $M_0$ as the action. This is possible, because following modification of the coset would not spoil the equations of motion:
\bea\label{newcoset}
{\tilde g} &=& e^{t\big( L_{-1} + \omega^2 L_1  \big)}e^{u \cdot G} e^{\Lambda\cdot M} = g e^{\Lambda\cdot M}  \nn \\
{\tilde g}^{-1}d{\tilde g} &=& e^{-\Lambda\cdot M}\wedge g^{-1}dg + e^{-\Lambda\cdot M}de^{\Lambda\cdot M}.
\eea
As the second term in ${\tilde g}^{-1}d{\tilde g} $ does not contain $G_{r,\mu}$ generators, and the adjoint action of the exponent is invertible, the conditions ${\widetilde \Omega}^{r,\mu}=0$ are equivalent to ${\Omega}^{r,\mu}=0$. Moreover, as $M$ forms in $g^{-1}dg$ vanish on these equations, the equations of motion for $\Lambda^A$ come only from the second term in ${\tilde g}^{-1}d{\tilde g}$, and, therefore, read just as ${\dot\Lambda}^A =0$.\footnote{The equations of motion for $u^{r,\mu}$ do not change if the exponents of $M$ generators are placed after the exponents of $G$, regardless of how exactly the $G$'s are arranged. In this case, the equations of motion for $\Lambda$ can be different, however. } Let us note that this only happens if the ${\Omega}^{r,\mu}=0$ forms vanish on the $u^{r,\mu}$ equations of motion, which was not the case of the paper \cite{KLS2}.

Let us mention some properties of the action constructed this way.
\begin{itemize}
\item The usual inverse Higgs effect \cite{IO1} is typically absent in such systems, as its presence in the case of spontaneously broken spacetime symmetries \cite{nonlin} actually imposes strong conditions on the structure constants of the algebra \cite{Goldst}. Though it is possible to put covariantly to zero not all $\Omega^{r,\mu}$ forms, they in general contain differentials of all variables. This means that resulting equations contain both fields and their derivatives, and equations cannot be solved algebraically for fields representing some of them in terms of others. (This does not preclude solving the whole set of equations for \textit{derivatives} of fields).
\item The action, constructed in the described way, contains derivatives of all variables, and is necessarily linear in them. It should be considered, therefore, in the Hamiltonian sense, with $u^{r,\mu}$ variables (parameters of $G_{r,\mu}$) representing both coordinates and momenta and $\Lambda^A$ variables being auxiliary to ensure invariance.
\item Proposed symmetry of systems of oscillators is dynamical, it cannot be realized on coordinates only.
\item The main variables $u^{r,\mu}$ parameterize the coset space
$$
\frac{{\cal G}}{SO(1,2)\times {\cal M}}.
$$
The action is made of connections on this space, which are related to the algebra ${\cal M}$.
\end{itemize}

\subsection{The equations of motion from the action}
It would be natural to expect that due to invariance of the action and the equations of motion, the necessary equations follow from the proposed action. Indeed, let us consider variation of this action
\bea\label{actvar}
S = \int {\widetilde\Omega}_{M_0} = \int {\tilde g}^{-1}d{\tilde g}|_{M_0}, \;\; \delta S = \int \big( {\tilde g}^{-1}d{\delta\tilde g}- {\tilde g}^{-1} \delta {\tilde g}  {\tilde g}^{-1}d{\tilde g}\big)|_{M_0} = \nn \\
= \int \big({\tilde g}^{-1}d{\tilde g}\,{\tilde g}^{-1}{\delta\tilde g}- {\tilde g}^{-1} \delta {\tilde g} \, {\tilde g}^{-1}d{\tilde g}\big)|_{M_0}.
\eea
One may notice that this variation is an integral of the commutator of the Cartan form of the extended coset and the forms, where differentials $du^{r,\mu}$, $d\lambda^A$ are replaced by the variations $\delta u^{r,\mu}$, $\delta\lambda^A$ (and $\delta t =0$, as $t$ is not varied). Therefore, the coefficients of $\delta u^{r,\mu}$, $\delta\lambda^A$ can only be combinations of expected equations of motion for $u^{r,\mu}$ and $\Lambda^A$. Some additional analysis would be required to find whether all of them are reproduced, and the $u(1)$ generator should be chosen carefully. In particular, if the ${\cal M}$ algebra is $so(1,2)$, $[M_a, M_b]=(a-b)M_{a+b}$, $a=-1,0,1$, it is required to put $M_{\pm 1}$ generators in the coset space, while the action will be obtained as the integral of the form of $M_0$. This will be illustrated by the following examples.

\section{Two harmonic oscillators and $so(2,3)$ algebra}
To make the statements above more clear, let us consider the example that can be treated in the homogenous parametrization completely. This is the system of two harmonic oscillators with identical frequency, which provide a realization of the  $so(2,3)$ algebra (already considered in \cite{KLS2}, but in different parametrization). The commutation relations of this algebra read
\bea\label{so23st}
&&\big[ L_m, L_n \big] = (m-n)L_{m+n}, \; m,n=-1,0,1; \big[ M_a, M_b \big] = (a-b)M_{a+b}, \; a,b=-1,0,1, \nn \\
&&\big[ L_m, G_{r,\mu}    \big] = \big( \frac{1}{2}m-r \big)G_{m+r,\mu}, \; r=-\frac{1}{2}, \frac{1}{2}, \; \; \big[ M_a, G_{r,\mu}    \big] = \big( \frac{1}{2}a-\mu \big)G_{m,a+\mu}, \; \; \mu=-\frac{1}{2}, \frac{1}{2}, \nn \\
&&\big[ G_{r,\mu},G_{s,\nu}  \big] =2\big( r\delta_{r+s,0}M_{\mu+\nu} + \mu \delta_{\mu+\nu,0}L_{r+s} \big).
\eea
It is evident that it is 5-graded in this basis (a basis with 3-grading also exists, but we do not consider it here). For our purposes, it would be convenient to rewrite them in the $SO(1,2)$ spinor notation, with $i,j,k,l=1,2$ and $\alpha,\beta,\mu,\nu=1,2 $:
\bea\label{so23alg}
&&[ L_{ij}, L_{kl} ] = \epsilon_{ik}L_{jl} + \epsilon_{jk}L_{il}+ \epsilon_{il}L_{jk}+ \epsilon_{jl}L_{ik}, \nn \\
&&[ M_{\alpha\beta}, M_{\mu\nu} ] = \epsilon_{\alpha\mu}M_{\beta\nu}+ \epsilon_{\beta\mu}M_{\alpha\nu} + \epsilon_{\alpha\nu}M_{\beta\mu}+ \epsilon_{\beta\nu}M_{\alpha\mu}, \nn \\
&&[ L_{ij},G_{k\alpha} ] = \epsilon_{ik}G_{j\alpha}+\epsilon_{jk}G_{i\alpha}, \;\; [M_{\alpha\beta}, G_{i\gamma}] = \epsilon_{\alpha\gamma}G_{i\beta} + \epsilon_{\beta\gamma}G_{i\alpha}, \\
&&[ G_{i\alpha}, G_{j\beta}] = - \epsilon_{ij}M_{\alpha\beta} - \epsilon_{\alpha\beta}L_{ij}. \nn
\eea
Here, $\epsilon_{ij}=-\epsilon_{ji}$, $\epsilon_{\alpha\beta}=-\epsilon_{\beta\alpha}$, $\epsilon_{12}=1$, $\epsilon^{ij}\epsilon_{jk}=\delta^i_k$,  $\epsilon^{\alpha\beta}\epsilon_{\beta\gamma}=\delta^\alpha_\gamma$.

The coset space can be parameterized as
\bea\label{so23coset}
g = e^{- \frac{1}{2} t \big( L_{11} + \omega^2 L_{22}  \big)}e^{u^{k\alpha}G_{k\alpha}}.
\eea
One may introduce the standard stereographical coordinates on $S^{2,2}$
\be\label{so23uz}
z^{i\alpha} = \frac{\tanh{\sqrt{\frac{u\cdot u}{2}}}}{\sqrt{\frac{u\cdot u}{2}}}u^{i\alpha}, \; u\cdot u = u^{i\alpha}u_{i\alpha}, \; z\cdot z = z^{i\alpha}z_{i\alpha}.
\ee
and write down the Cartan forms of interest as
\bea\label{so23CF}
&\big(\Omega_{G} \big)^{1\alpha}= \frac{dz^{1\alpha}-dt z^{2\alpha}}{1-\frac{1}{2}z\cdot z}, \; \big(\Omega_{G} \big)^{2\alpha}= \frac{dz^{2\alpha}+dt \omega^2 z^{1\alpha}}{1-\frac{1}{2}z\cdot z},& \nn \\
&\big(\Omega_M \big)^{\alpha\beta} = \frac{1}{2}\frac{ z^{i(\alpha}dz^{\beta)}_i + dt \big( z^{2\alpha} z^{2\beta} +\omega^2 z^{1\alpha} z^{1\beta}\big)  }{1 - \frac{1}{2}z\cdot z}.&
\eea
From the conditions $\big(\Omega_{G} \big)^{i\alpha}=0$ one may obtain the equations
\be\label{so23osc2}
\dot{z}{}^{1\alpha} = z^{2\alpha}, \; \dot{z}{}^{2\alpha} = -\omega^2 z^{1\alpha} \; \Rightarrow {\ddot z}{}^{1\alpha} + \omega^2 z^{1\alpha}=0,
\ee
which are the equations of motion of two harmonic oscillators. Let us note that $u^{i\alpha}$ satisfy similar equations, as $u\cdot u$ and $z\cdot z$ are the constants of motion.

To construct the action, one should extend the coset space by two more generators
\be\label{so23cosetext}
{\tilde g} = g \, e^{\Lambda^{11}M_{11} + \Lambda^{22}M_{22}}, \; \; \lambda^{\alpha\beta}=\Lambda^{\alpha\beta}\frac{\tan{2\sqrt{\Lambda^{11}\Lambda^{22}}}}{2\sqrt{\Lambda^{11}\Lambda^{22}}}.
\ee
The action is an integral of the $\big({\widetilde\Omega}_M \big)^{12}$ Cartan form of this extended coset
\be\label{so23lagr}
S= \int \Big\{ \frac{\lambda^{22}d\lambda^{11} -\lambda^{11}d\lambda^{22} }{1+ 4 \lambda^{11}\lambda^{22}}  + \big(\Omega_M \big)^{12} \frac{1 - 4 \lambda^{11}\lambda^{22}}{1+ 4 \lambda^{11}\lambda^{22}}  + 2\frac{\lambda^{11}\big(\Omega_M \big)^{22} -\lambda^{22} \big(\Omega_M \big)^{11}}{1+ 4 \lambda^{11}\lambda^{22}}\Big\}
\ee
It can be directly checked that the equations of motion \p{so23osc2} as well as ${\dot\lambda}^{\alpha\beta} =0$ follow from this action. The equations ${\dot\lambda}^{\alpha\beta} =0$ also follow from the conditions $\big({\widetilde\Omega}_M \big)^{11}=0$, $\big({\widetilde\Omega}_M \big)^{22}=0$ on the new Cartan forms
\bea\label{so23extCF}
\big({\widetilde\Omega}_M \big)^{11} = \frac{d\lambda^{11}}{1+4  \lambda^{11}\lambda^{22} } + \frac{\big(\Omega_M \big)^{11} - 4 \lambda^{11}\big(\Omega_M \big)^{12} + 4 (\lambda^{11})^2 \big(\Omega_M \big)^{22}}{1+4  \lambda^{11}\lambda^{22}}, \nn \\
\big({\widetilde\Omega}_M \big)^{22} = \frac{d\lambda^{22}}{1+4  \lambda^{11}\lambda^{22} } + \frac{\big(\Omega_M \big)^{22} + 4 \lambda^{22}\big(\Omega_M \big)^{12} + 4 (\lambda^{22})^2 \big(\Omega_M \big)^{11}}{1+4  \lambda^{11}\lambda^{22}},
\eea
as expected.

The action \p{so23lagr} is of the first order and should be treated as the Hamiltonian one with the factor of $dt$ in \p{so23lagr} being the Hamiltonian itself and the rest being the source of the Poisson brackets. In describing the Hamiltonian and brackets, it is very useful to define set of functions
\bea\label{so23hfuncs}
&& h^{11} = \frac{4 \lambda^{11}}{(1+4  \lambda^{11}\lambda^{22}) \big(1 - \frac{z\cdot z}{2} \big)}, \; h^{22} = \frac{-4 \lambda^{22}}{(1+4  \lambda^{11}\lambda^{22}) \big(1 - \frac{z\cdot z}{2} \big)},\nn \\
&& h^{12}=h^{21} = \frac{-(1-4  \lambda^{11}\lambda^{22})}{(1+4  \lambda^{11}\lambda^{22}) \big(1 - \frac{z\cdot z}{2} \big)},
\eea
which the action \p{so23lagr} is composed of. Then the action can be rewritten as
\be\label{so23lagr2}
S= \int \Big\{ \frac{\lambda^{22}d\lambda^{11} -\lambda^{11}d\lambda^{22} }{1+ 4 \lambda^{11}\lambda^{22}}  + \frac{1}{4}h_{\alpha\beta}\,z^{i\alpha}dz_{i}^\beta -dt H\Big\}, \;\;
H = -\frac{1}{2} z^{2\alpha}z^{2\beta}h_{\alpha\beta} - \frac{\omega^2}{2}  z^{1\alpha}z^{1\beta}h_{\alpha\beta}.
\ee
Differential of the first part of this action provides the brackets between variables. The brackets between $\lambda^{11}$ and $\lambda^{22}$, as well as between $z^{i\alpha}$ are reasonably simple
\be\label{so23bracs1}
\big\{ \lambda^{11}, \lambda^{22}  \big\} = - \frac{1}{4}\Big( 1 - \frac{z\cdot z}{2}  \Big)\big( 1+ 4\lambda^{11}\lambda^{22}  \big)^2, \; \; \big\{ z^{i\alpha},z^{j\beta}  \big\} = \Big( 1 - \frac{z\cdot z}{2}  \Big)^3 \;\epsilon^{ij}\,h^{\alpha\beta}.
\ee
The brackets between $\lambda^{\alpha\beta}$ and $z^{i\gamma}$ are not that systematic. They, however, imply relatively simple brackets of $z^{i\alpha}$ with the functions $h^{\mu\nu}$, as well as between $h^{\alpha\beta}$ and $h^{\mu\nu}$:
\be\label{so23bracs3}
\big\{ h^{\alpha\beta}, h^{\mu\nu}   \big\} = \epsilon^{\alpha\mu}h^{\beta\nu} +  \epsilon^{\beta\mu}h^{\alpha\nu}+\epsilon^{\alpha\nu}h^{\beta\mu} +  \epsilon^{\beta\nu}h^{\alpha\mu}, \;\;
 \big\{ h^{\alpha\beta}, z^{i\gamma} \big\} = \epsilon^{\alpha\gamma}z^{i\beta} + \epsilon^{\beta\gamma}z^{i\alpha}.
\ee
As expected, these brackets and Hamiltonian \p{so23lagr2} reproduce the expected equations of motion \p{so23osc2}, along with $\dot{\lambda}^{11}=0$, $\dot{\lambda}^{22}=0$. One may finally note that the following quantities reproduce the $so(2,3)$ algebra with respect to the Poisson brackets:
\be\label{so23algbracs}
{\widetilde L}_{ij} = z_{i\alpha}z_{j\beta}h^{\alpha\beta}, \;\; {\widetilde M}_{\alpha\beta} =-h_{\alpha\beta}, \; \; {\widetilde G}_{i\alpha} = z_i^{\beta}h_{\alpha\beta}.
\ee

\section{$G_{2(2)}$ oscillators}
Let us note that the considered homogeneous parametrization is not only one that allows to obtain linear equations of motion. For example, let us consider the exceptional $G_{2(2)}$ algebra. Commutation relations of its 14 generators read
\bea\label{G2alg1}
&&\big[ L_{m},L_{n}   \big]= (m-n)L_{m+n},\; m,n=-1,0,1,\; \big[ M_{a},M_{b}   \big]= (a-b)M_{a+b}, \;a,b =-1,0,1, \nn \\
&&\big[ L_{m}, G_{r,\mu} \big] = \Big( \frac{m}{2}-r\Big)G_{r+m,\mu}, \; \big[ M_{a}, G_{r,\mu} \big] = \Big( \frac{3a}{2}-r  \Big)G_{r,\mu+a}, \; r=-\frac{1}{2},\frac{1}{2},\, \mu= -\frac{3}{2},-\frac{1}{2},\frac{1}{2},\frac{3}{2}, \nn \\
&&\big[ G_{r,\mu}, G_{s,\nu}   \big] = \, r \, \delta_{r+s,0} \big( 6 \mu^2 + 6\nu^2 - 8 \mu\nu -9  \big) M_{\mu+\nu}+ 3 \, \mu \big( 4\mu^2-5\big) \delta_{\mu+\nu, 0} L_{r+s}.
\eea
Looking at the commutators \p{G2alg1}, one may note that this algebra contains two different $sl(2)$ algebras and can be viewed as $5$- or $7$-graded, depending on whether $L$ or $M$ algebra is considered as the primary one. We may, therefore, consider two coset spaces
\bea\label{G2cosetmod1}
\hat{g}_1 &=& e^{t (L_{-1}+\omega^2 L_1)} \prod_{\mu = -3/2}^{3/2} e^{u_\mu G_{-1/2,\mu}+v_\mu G_{1/2,\mu}},  \\
\label{G2cosetmod2}
\hat{g}_2 &=& e^{t (M_{-1}+\omega^2 M_1)} e^{\sum_{\mu=-3/2}^{3/2} x_\mu G_{-1/2,\mu}}\, e^{\sum_{\mu=-3/2}^{3/2} y_{\mu}G_{1/2,\mu}}.
\eea
The generators $G$ in these parameterizations are arranged in such a way that 1) generators inside each exponent form closed subsets under the action of the conformal generators, 2) exponents themselves are placed so that second subset of generators can be produced from the first by the action of $M_{1}$ (or $L_{1}$), and so on. For this reason, this structure of the coset can be called ``matryoshka''. It can be checked that in both cases the equations of motion are again linear.
\subsection{Four harmonic oscillators}
Let us briefly consider the coset space $\hat{g}_1$ \p{G2cosetmod1}. Evaluating its Cartan forms, one may note that all forms of $G_{r,\mu}$ are composed of subforms
\be\label{hatg1subf}
U_\mu = du_\mu - v_\mu\,  dt, \; V_\mu = dv_\mu + \omega^2 u_\mu\,  dt.
\ee
Therefore, the equations $\Omega_{r,\mu}=0$ are the linear system of equations w.r.t. $U_\mu$, $V_\mu$ with zero right-hand side. As this system appears to be non-degenerate, it implies
\be\label{4osceom}
U_\mu = V_\mu=0 \; \Rightarrow \; \ddot u_{\mu} + \omega^2 u_{\mu} =0,
\ee
which are just the equations of motion of four harmonic oscillators with the same frequency. The system of only four equations $\Omega_{-1/2,\mu}=0$ cannot be solved for $v_\mu$, and the inverse Higgs effect is also absent in this basis.

The action for this system can also be constructed by extending the coset space $\hat{g}_1 \rightarrow \tilde{\hat g}_1 =\hat{g}_1 \exp(\Lambda_{-1}L_{-1} + \Lambda_1 L_1)$. It reads
\be\label{hatg1act}
S = \int {\widetilde\Omega}_{M_0} = \int \Big[ \frac{\lambda_{-1}d\lambda_1 - \lambda_1 d\lambda_{-1}}{1+ \lambda_1 \lambda_{-1}}+ \frac{\big( 1 -  \lambda_1 \lambda_{-1}  \big)\Omega_{M_0} + 2\lambda_{-1}\Omega_{M_1}- 2\lambda_{1}\Omega_{M_{-1}} }{1+ \lambda_1 \lambda_{-1}} \Big],
\ee
where we defined
\be\label{lambdaredefhat}
\lambda_{\pm 1} = \Lambda_{\pm 1} \frac{\tan{\sqrt{\Lambda_{-1}\Lambda_1}}}{\sqrt{\Lambda_{-1}\Lambda_1}}
\ee
and $\Omega_{M_a}$ are the $M_a$ forms of the $\hat{g}_1$ coset. It can be checked explicitly that the action \p{hatg1act} produces the expected equations of motion $\dot u_\mu = v_\mu$, $\dot v_\mu =- \omega^2 u_\mu$, $\dot\lambda_{\pm1}=0$.

\subsection{Two Pais-Uhlenbeck oscillators}
The coset ${\hat g}_2$ \p{G2cosetmod2} also leads to linear equations. The Cartan forms in this case are composed of subforms
\bea\label{hatg2subf}
&&X_{-3/2} = dx_{-3/2}-x_{1/2}dt, \; \; X_{-1/2} = dx_{-1/2} -2x_{1/2}dt + 3 \omega^2 x_{3/2}dt, \nn \\
&&X_{1/2} = dx_{1/2} - 3 x_{3/2}dt + 2 \omega^2 x_{-1/2}dt, \;\; X_{3/2} = dx_{3/2} +\omega^2 x_{1/2}dt,  \\
&&Y_{-3/2} = dy_{-3/2}-y_{1/2}dt, \; \; Y_{-1/2} = dy_{-1/2} -2y_{1/2}d\tau + 3 \omega^2 y_{3/2}dt,\nn \\
&&Y_{1/2} = dy_{1/2} - 3 y_{3/2}dt + 2 \omega^2 y_{-1/2}dt, \;\; Y_{3/2} = dy_{3/2} +\omega^2 y_{1/2}dt. \nn
\eea
Here, it makes sense to introduce completely symmetric 3-spinor notation by the relation
\be
x_{-3/2} = \frac{1}{3} x^{1,1,1}, \; x_{-1/2} = x^{1,1,2}, \; x_{1/2} = x^{1,2,2}, \; x_{3/2}  = \frac{1}{3}x^{2,2,2}.
\ee
As a result, the $G_{r,\mu}$ and $L_m$ Cartan forms can be written as
\bea\label{hatg2GLforms}
&\Omega_{L_{-1}} = xX, \; \Omega_{L_0} = 2 yX,\; \Omega_{L_1} = yY +2 (y^3X) - \frac{1}{2}y^4\cdot xX,& \nn \\
&\Omega_{-1/2}^{\alpha\beta\gamma} = X^{\alpha\beta\gamma} + y^{\alpha\beta\gamma} \cdot xX, \; \Omega_{1/2}^{\alpha\beta\gamma} = Y^{\alpha\beta\gamma} -3 (yXy)^{\alpha\beta\gamma}+ y^{\alpha\beta\gamma}(yX) + \big(y^3 \big)^{\alpha\beta\gamma}\cdot xX.&
\eea
Here, we used the notation
\bea\label{combs}
&&xy = x^{\alpha\beta\gamma}y_{\alpha\beta\gamma},\;\; y^4 =y^{\alpha\mu\nu}y_{\mu\nu\sigma}y^{\sigma\beta\gamma}y_{\alpha\beta\gamma},  \\
&&(xyz)^{\alpha\beta\gamma} = \frac{1}{3}\big( x^{\alpha\mu\nu}y_{\mu\nu\sigma}z^{\sigma\beta\gamma}+ x^{\beta\mu\nu}y_{\mu\nu\sigma}z^{\sigma\alpha\gamma}+x^{\gamma\mu\nu}y_{\mu\nu\sigma}z^{\sigma\alpha\beta}  \big). \nn
\eea
It is obvious to show that the equations $\Omega_{r}^{\alpha\beta\gamma}= 0$ imply  $X^{\alpha\beta\gamma}=0$ and $Y^{\alpha\beta\gamma}=0$. Excluding from them all variables in terms of derivatives of $x_{-3/2}$ and $y_{-3/2}$, one may find
\be\label{hatg2eoms}
\Big(\frac{d}{dt^2} + (3\omega)^2  \Big)\Big(\frac{d}{dt^2} + \omega^2  \Big)x_{-3/2}=0, \;\; \Big(\frac{d}{dt^2} + (3\omega)^2  \Big)\Big(\frac{d}{dt^2} + \omega^2  \Big)y_{-3/2}=0
\ee
-the equations of motion of two Pais-Uhlenbeck oscillators with frequencies $\omega$ and $3\omega$.

One may check explicitly that the action
\be\label{hatg2act}
S = \int {\widetilde\Omega}_{L_0} = \int \Big[ \frac{\lambda_{-1} d\lambda_1- \lambda_1 d\lambda_{-1}}{1 + \lambda_{-1} \lambda_1}  + \frac{\big( 1 - \lambda_{-1} \lambda_1  \big)\Omega_{L_0} + 2 \lambda_{-1}\Omega_{L_1} -2 \lambda_{1} \Omega_{L_{-1}}}{1 + \lambda_{-1} \lambda_1}     \Big]
\ee
reproduces the equations of motion \p{hatg2eoms}. As before, it can be obtained by extending the coset as $\hat{g}_2 \rightarrow \tilde{\hat g}_2 =\hat{g}_2 \exp(\Lambda_{-1}L_{-1} + \Lambda_1 L_1)$, evaluating the modified $L_0$ Cartan form and performing substitution \p{lambdaredefhat}.

\section{Conclusion}
In this paper, we demonstrated that for any noncompact simple Lie algebras one may construct a dynamical realization, then the conditions on the Cartan forms of the specifically chosen coset space reproduce the equations of motion of the harmonic and/or Pais-Uhlenbeck oscillators. Also, we showed that for each appearing system of oscillators, one may provide the first-order action, which is invariant with respect to the whole group and produces the expected equations of motion.

The studied examples show that one may parameterize the coset space in a few different ways and still obtain the linear equations of motion. One way was used in general proof and can be called ``homogeneous parametrization'', as it involves placing of all dynamical variables and their generators into just a single exponent. Another parametrization, which can be called ``matryoshka'', was found for the $G_{2(2)}$ algebra and involves placing generators into the exponents according to their properties with respect to the conformal and automorphism $sl(2)$ algebras. Let us note that such a parametrization is also known for the $so(2,3)$ algebra, and it is possible that it is another universal parametrization which ensures the linearity of the equations of motion.

Each of these two parameterizations has its own advantages and disadvantages. The homogeneous parametrization is suitable for general proofs, can be constructed for any group, the fields in it are in linear representation of conformal and automorphism groups, which provides a more natural connection to the geometry. However, the calculations in this parametrization involve infinite matrix power series, which sometimes are very difficult to evaluate explicitly. Contrary to this, in the ``matryoshka'' parametrization the Cartan forms involve the finite number of terms, which are always possible to find explicitly. The variables involved, however, do not form a linear representation of the automorphism group, which can make the Cartan forms longer, complicated and difficult to deal with.

A typical feature of the considered parameterizations is the absence of the usual inverse Higgs effect, which necessitates the related actions to be considered  as Hamiltonian ones, with $u^{r,\mu}$ variables representing both coordinates and momenta. The symmetry then will also be dynamical, which is impossible to realize off-shell in terms of coordinates only.

\section*{Acknowledgments}
The work of N.K. and S.K. was supported by the RSCF, grant 14-11-00598.


\section*{References}
\begin{thebibliography}{99}
\bibitem{KN1}
Krivonos S and Nersessian A 2017 SU(1,2) invariance in two-dimensional oscillator {\it JHEP} {\bf 1702} 006 ({\it Preprint} arXiv:1610.02499)

\bibitem{KLS2}
Krivonos S, Lechtenfeld O and Sorin A 2017 Hidden symmetries of deformed oscillators {\it Nucl. Phys. } {\bf B924} 33-46 ({\it Preprint} arXiv:1612.07832)

\bibitem{phenoml1}
Coleman S, Wess J and Zumino B 1969 Structure of phenomenological lagrangians. 1 {\it Phys. Rev.} {\bf 177} 2239,

\bibitem{phenoml2}
Callan C, Coleman S, Wess J and Zumino B 1969 Structure of phenomenological lagrangians. 2 {\it Phys. Rev.} {\bf 177} 2247

\bibitem{phenoml3}
Volkov D 1973 Phenomenological lagrangians {\it Sov. J. Part. Nucl.} {\bf 4} 3

\bibitem{nonlin}
Ogievetsky V 1974 Nonlinear realizations of internal and spacetime symmetries {\it Acta Universitatis Wratislaviensis } {\bf 207} 117 (Proc. of the X'th Winter school of theoretical physics in Karpacz)

\bibitem{BG}
Bina B and G\"{u}nadin M 1997 Real forms of nonlinear superconformal and quasi-superconformal algebras and their unified realization {\it Nucl. Phys.} {\bf B502} 713 ({\it Preprint} hep-th/9703188)

\bibitem{KLS1}
Krivonos S, Lechtenfeld O and Sorin A 2016 Minimal realization of $\ell$-conformal Galilei algebra, Pais-Uhlenbeck oscillators and their deformation {\it JHEP} {\bf 1610} 078 ({\it Preprint} arXiv:1607.03756)

\bibitem{Galaj1}
Galajinsky A 2010 Conformal mechanics in Newton-Hooke spacetime {\it Nucl. Phys.}  {\bf B832} 586 ({\it Preprint} arXiv:1002.2290)

\bibitem{AGGM}
Andrzejewski K, Galajinsky A, Gonera J and Masterov I 2014 Conformal Newton-Hooke symmetry of Pais-Uhlenbeck oscillator {\it Nucl. Phys. } {\bf B885} 150-162 ({\it Preprint} arXiv:1402.1297)

\bibitem{IO1}
Ivanov E and Ogievetsky V 1975 The Inverse Higgs Phenomenon in Nonlinear Realizations {\it Teor.Mat.Phys} {\bf 25} 164-177

\bibitem{Goldst}
Klein R, Roest D and Stefanyszyn D 2017  Spontaneously Broken Spacetime Symmetries and the Role of Inessential Goldstones ({\it Preprint} arXiv:1709.03525)
\end{thebibliography}
\end{document}